\documentclass[twocolumn,preprintnumbers,pra,hyperref]{revtex4}
\usepackage{amssymb}
\usepackage{amsfonts}
\usepackage{amsmath}
\usepackage{graphicx}

\begin{document}

\title{Evolution of the single-mode squeezed vacuum state in amplitude
dissipative channel }
\author{Hong-Yi Fan$^{1}$, Shuai Wang$^{1}$ and Li-Yun Hu$^{2\ast }$\thanks{
{\small E-mail: hlyun2008@126.com.}}}
\affiliation{$^{1}${\small Department of Physics, Shanghai Jiao Tong University, Shanghai
200240,China}\\
$^{2}${\small College of Physics \& Communication Electronics, Jiangxi
Normal University, Nanchang 330022, China}\\
$\ast ${\small Corresponding author.} {\small E-mail: hlyun2008@126.com.}}

\begin{abstract}
{\small Using the way of deriving infinitive sum representation of density
operator as a solution to the master equation describing the amplitude
dissipative channel by virtue of the entangled state representation, we show
manifestly how the initial density operator of a single-mode squeezed vacuum
state evolves into a definite mixed state which turns out to be a squeezed
chaotic state with decreasing-squeezing. We investigate average photon
number, photon statistics distributions for this state.}
\end{abstract}

\maketitle

\section{Introduction}

Squeezed states are such for which the noise in one of the chosen pair of
observables is reduced below the vacuum or ground-state noise level, at the
expense of increased noise in the other observable. The squeezing effect
indeed improves interferometric and spectroscopic measurements, so in the
context of interferometric detection and of gravitational waves the squeezed
state is very useful \cite{01,02}. In a very recently published paper,
Agarwal \cite{r1} revealed that a vortex state of a two-mode system can be
generated from a squeezed vacuum by subtracting a photon, such a subtracting
mechanism may happen in a quantum channel with amplitude damping. Usually,
in nature every system is not isolated, dissipation or dephasing usually
happens when a system is immersed in a thermal environment, or a signal (a
quantum state) passes through a quantum channel which is described by a
master equation \cite{03}. For example, when a pure state propagates in a
medium, it inevitably interacts with it and evolves into a mixed state \cite%
{04}. Dissipation or dephasing will deteriorate the degree of
nonclassicality of photon fields, so physicists pay much attention to it
\cite{05,06,07}.

In this present work we investigate how an initial single-mode squeezed
vacuum state evolves in an amplitude dissipative channel (ADC). When a
system is described by its interaction with a channel with a large number of
degrees of freedom, master equations are set up for a better understanding
how quantum decoherence is processed to affect unitary character in the
dissipation or gain of the system. In most cases people are interested in
the evolution of the variables associated with the system only. This
requires us to obtain the equations of motion for the system of interest
only after tracing over the reservoir variables. A quantitative measure of
nonclassicality of quantum fields is necessary for further investigating the
system's dynamical behavior. For this channel, the associated loss mechanism
in physical processes is governed by the following master equation \cite{03}
\begin{equation}
\frac{d\rho \left( t\right) }{dt}=\kappa \left( 2a\rho a^{\dagger
}-a^{\dagger }a\rho -\rho a^{\dagger }a\right) ,  \label{p1}
\end{equation}
where $\rho $ is the density operator of the system, and $\kappa $ is the
rate of decay. We have solved this problem with use of the thermo entangled
state representation \cite{08}. Our questions are: What kind of mixed state
does the initial squeezed state turns into? How does the photon statistics
distributions varies in the ADC?

Thus solving master equations is one of the fundamental tasks in quantum
optics. Usually people use various quasi-probability representations, such
as P-representation, Q-representation, complex P-representation, and Wigner
functions, etc. for converting the master equations of density operators
into their corresponding c-number equations. Recently, a new approach \cite
{08,09}, using the thermal entangled state representation \cite{10,11} to
convert operator master equations to their c-number equations is presented
which can directly lead to the corresponding Kraus operators (the infinitive
representation of evolved density operators) in many cases.

The work is arranged as follows. In Sec. 2 by virtue of the entangled state
representation we briefly review our way of deriving the infinitive sum
representation of density operator as a solution of the master equation. In
Sec. 3 we show that a pure squeezed vacuum state (with squeezing parameter $
\lambda )$ will evolves into a mixed state (output state), whose exact form
is derived, which turns out to be a squeezed chaotic state. We investigate
average photon number, photon statistics distributions for this state. The
probability of finding $n$ photons in this mixed state is obtained which
turns out to be a Legendre polynomial function relating to the squeezing
parameter $\lambda $ and the decaying rate $\kappa $. In Sec. 4 we discuss
the photon statistics distributions of the output state. In Sec. 5 and 6 we
respectively discuss the Wigner function and tomogram of the output state.

\section{Brief review of deducing the infinitive sum representation of $
\protect\rho \left( t\right) $}

For solving the above master equation, in a recent review paper \cite{12} we
have introduced a convenient approach in which the two-mode entangled state
\cite{10,11}
\begin{equation}
|\eta \rangle =\exp (-\frac{1}{2}|\eta |^{2}+\eta a^{\dag }-\eta ^{\ast }
\tilde{a}^{\dag }+a^{\dag }\tilde{a}^{\dag })|0\tilde{0}\rangle ,  \label{p2}
\end{equation}
is employed, where $\tilde{a}^{\dag }$ is a fictitious mode independent of
the real mode $a^{\dagger },$ $[\tilde{a},a^{\dagger }]=0$. $|\eta =0\rangle
$ possesses the properties
\begin{align}
a|\eta & =0\rangle =\tilde{a}^{\dag }|\eta =0\rangle ,  \notag \\
a^{\dag }|\eta & =0\rangle =\tilde{a}|\eta =0\rangle ,  \label{p3} \\
(a^{\dag }a)^{n}|\eta & =0\rangle =(\tilde{a}^{\dag }\tilde{a})^{n}|\eta
=0\rangle .  \notag
\end{align}
Acting the both sides of Eq.(\ref{p1}) on the state $|\eta =0\rangle \equiv
\left \vert I\right \rangle $, and denoting $\left \vert \rho \right \rangle
=\rho \left \vert I\right \rangle $, we have
\begin{align}
\frac{d}{dt}\left \vert \rho \right \rangle & =\kappa \left( 2a\rho
a^{\dagger }-a^{\dagger }a\rho -\rho a^{\dagger }a\right) \left \vert
I\right \rangle  \notag \\
& =\kappa \left( 2a\tilde{a}-a^{\dagger }a-\tilde{a}^{\dagger }\tilde{a}
\right) \left \vert \rho \right \rangle ,  \label{p4}
\end{align}
so its formal solution is
\begin{equation}
\left \vert \rho \right \rangle =\exp \left[ \kappa t\left( 2a\tilde{a}
-a^{\dagger }a-\tilde{a}^{\dagger }\tilde{a}\right) \right] \left \vert \rho
_{0}\right \rangle ,  \label{p5}
\end{equation}
where $\left \vert \rho _{0}\right \rangle \equiv \rho _{0}\left \vert
I\right \rangle ,$ $\rho _{0}$ is the initial density operator.

Noticing that the operators in Eq.(\ref{p5}) obey the following commutative
relation,
\begin{equation}
\left[ a\tilde{a},a^{\dagger }a\right] =\left[ a\tilde{a},\tilde{a}^{\dagger
}\tilde{a}\right] =\tilde{a}a  \label{p6}
\end{equation}
and%
\begin{equation}
\left[ \frac{a^{\dagger }a+\tilde{a}^{\dagger }\tilde{a}}{2},a\tilde{a}
\right] =-\tilde{a}a,  \label{p7}
\end{equation}
as well as using the operator identity \cite{13}
\begin{equation}
e^{\lambda \left( A+\sigma B\right) }=e^{\lambda A}e^{\sigma \left(
1-e^{-\lambda \tau }\right) B/\tau },  \label{p8}
\end{equation}%
(which is valid for $\left[ A,B\right] =\tau B$), we have
\begin{equation}
e^{-2\kappa t\left( \frac{a^{\dagger }a+\tilde{a}^{\dagger }\tilde{a}}{2}-a
\tilde{a}\right) }=e^{-\kappa t\left( a^{\dagger }a+\tilde{a}^{\dagger }
\tilde{a}\right) }e^{T^{\prime }a\tilde{a}},  \label{p9}
\end{equation}
where $T^{\prime }=1-e^{-2\kappa t}.$

Then substituting Eq.(\ref{p9}) into Eq.(\ref{p5}) yields \cite{12}
\begin{align}
\left \vert \rho \right \rangle & =e^{-\kappa t\left( a^{\dagger }a+\tilde{a}
^{\dagger }\tilde{a}\right) }\sum_{n=0}^{\infty }\frac{T^{\prime n}}{n!}
a^{n} \tilde{a}^{n}\left \vert \rho _{0}\right \rangle  \notag \\
& =e^{-\kappa ta^{\dagger }a}\sum_{n=0}^{\infty }\frac{T^{\prime n}}{n!}
a^{n}\rho _{0}a^{\dag n}e^{-\kappa t\tilde{a}^{\dagger }\tilde{a}}\left
\vert I\right \rangle  \notag \\
& =\sum_{n=0}^{\infty }\frac{T^{\prime n}}{n!}e^{-\kappa ta^{\dagger
}a}a^{n}\rho _{0}a^{\dag n}e^{-\kappa ta^{\dagger }a}\left \vert I\right
\rangle ,  \label{p10}
\end{align}%
which leads to the infinitive operator-sum representation of$\ \rho $,
\begin{equation}
\rho =\sum_{n=0}^{\infty }M_{n}\rho _{0}M_{n}^{\dagger },  \label{p11}
\end{equation}
where
\begin{equation}
M_{n}\equiv \sqrt{\frac{T^{\prime n}}{n!}}e^{-\kappa ta^{\dagger }a}a^{n}.
\label{p12}
\end{equation}
We can prove
\begin{align}
\sum_{n}M_{n}^{\dagger }M_{n}& =\sum_{n}\frac{T^{\prime n}}{n!}a^{\dag
n}e^{-2\kappa ta^{\dagger }a}a^{n}  \notag \\
& =\sum_{n}\frac{T^{\prime n}}{n!}e^{2n\kappa t}\colon a^{\dag n}a^{n}\colon
e^{-2\kappa ta^{\dagger }a}  \notag \\
& =\left. :e^{T^{\prime }e^{2\kappa t}a^{\dagger }a}:\right. e^{-2\kappa
ta^{\dagger }a}  \notag \\
& =\left. :e^{\left( e^{2\kappa t}-1\right) a^{\dagger }a}:\right.
e^{-2\kappa ta^{\dagger }a}=1,  \label{p13}
\end{align}
where $\colon \colon $ stands for the normal ordering. Thus $M_{n}$ is a
kind of Kraus operator, and $\rho $ in Eq.(\ref{p11}) is qualified to be a
density operator, i.e.,
\begin{equation}
Tr\left[ \rho \left( t\right) \right] =Tr\left[ \sum_{n=0}^{\infty
}M_{n}\rho _{0}M_{n}^{\dagger }\right] =Tr\rho _{0}.  \label{p14}
\end{equation}
Therefore, for any given initial state $\rho _{0}$, the density operator $
\rho \left( t\right) $ can be directly calculated from Eq.(\ref{p11}). The
entangled state representation provides us with an elegant way of deriving
the infinitive sum representation of density operator as a solution of the
master equation.

\section{Evolving of an initial single-mode squeezed vacuum state in ADC}

It is seen from Eq.(\ref{p11}) that for any given initial state $\rho _{0}$,
the density operator $\rho \left( t\right) $ can be directly calculated.
When $\rho _{0}$ is a single-mode squeezed vacuum state,
\begin{equation}
\rho _{0}=\text{sech}\lambda \exp \left( \frac{\tanh \lambda }{2}a^{\dag
2}\right) \left\vert 0\right\rangle \left\langle 0\right\vert \exp \left(
\frac{\tanh \lambda }{2}a^{2}\right) ,  \label{p15}
\end{equation}
we see
\begin{eqnarray}
\rho \left( t\right) &=&\text{sech}\lambda \sum_{n=0}^{\infty }\frac{
T^{\prime n}}{n!}e^{-\kappa ta^{\dagger }a}a^{n}\exp \left( \frac{\tanh
\lambda }{2}a^{\dag 2}\right) \left\vert 0\right\rangle  \notag \\
&&\times \left\langle 0\right\vert \exp \left( \frac{\tanh \lambda }{2}
a^{2}\right) a^{\dag n}e^{-\kappa ta^{\dagger }a}.  \label{p16}
\end{eqnarray}
Using the Baker-Hausdorff lemma \cite{14},
\begin{equation}
e^{\lambda \hat{A}}\hat{B}e^{-\lambda \hat{A}}=\hat{B}+\lambda \left[ \hat{A}
,\hat{B}\right] +\frac{\lambda ^{2}}{2!}\left[ \hat{A},\left[ \hat{A},\hat{B}
\right] \right] +\cdots .  \label{p17}
\end{equation}
we have
\begin{eqnarray}
a^{n}\exp \left( \frac{\tanh \lambda }{2}a^{\dag 2}\right) \left\vert
0\right\rangle &=&e^{\frac{\tanh \lambda }{2}a^{\dag 2}}e^{-\frac{\tanh
\lambda }{2}a^{\dag 2}}a^{n}e^{\frac{\tanh \lambda }{2}a^{\dag 2}}\left\vert
0\right\rangle  \notag \\
&=&e^{\frac{\tanh \lambda }{2}a^{\dag 2}}\left( a+a^{\dagger }\tanh \lambda
\right) ^{n}\left\vert 0\right\rangle .  \label{p18}
\end{eqnarray}
Further employing the operator identity \cite{15}
\begin{equation}
\left( \mu a+\nu a^{\dagger }\right) ^{m}=\left( -i\sqrt{\frac{\mu \nu }{2}}
\right) ^{m}\colon H_{m}\left( i\sqrt{\frac{\mu }{2\nu }}a+i\sqrt{\frac{\nu
}{2\mu }}a^{\dagger }\right) \colon ,  \label{p19}
\end{equation}
where $H_{m}(x)$ is the Hermite polynomial, we know
\begin{eqnarray}
&&\left( a+a^{\dagger }\tanh \lambda \right) ^{n}  \notag \\
&=&\left( -i\sqrt{\frac{\tanh \lambda }{2}}\right) ^{n}\colon H_{n}\left( i
\sqrt{\frac{1}{2\tanh \lambda }}a+i\sqrt{\frac{\tanh \lambda }{2}}a^{\dagger
}\right) \colon .  \label{p20}
\end{eqnarray}
From Eq.(\ref{p18}), it follows that
\begin{eqnarray}
a^{n}e^{\frac{\tanh \lambda }{2}a^{\dag 2}}\left\vert 0\right\rangle
&=&\left( -i\sqrt{\frac{\tanh \lambda }{2}}\right) ^{n}e^{\frac{\tanh
\lambda }{2}a^{\dag 2}}  \notag \\
&&\times H_{n}\left( i\sqrt{\frac{\tanh \lambda }{2}}a^{\dagger }\right)
\left\vert 0\right\rangle .  \label{p21}
\end{eqnarray}
On the other hand, noting $e^{-\kappa ta^{\dagger }a}a^{\dagger }e^{\kappa
ta^{\dagger }a}=a^{\dagger }e^{-\kappa t},e^{\kappa ta^{\dagger
}a}ae^{-\kappa ta^{\dagger }a}=ae^{-\kappa t}$ and the normally ordered form
of the vacuum projector $\left\vert 0\right\rangle \left\langle 0\right\vert
=\colon e^{-a^{\dagger }a}\colon ,$ we have
\begin{align}
\rho \left( t\right) & =\text{sech}\lambda \sum_{n=0}^{\infty }\frac{
T^{\prime n}}{n!}e^{-\kappa ta^{\dagger }a}a^{n}e^{\frac{\tanh \lambda }{2}
a^{\dag 2}}\left\vert 0\right\rangle  \notag \\
& \times \left\langle 0\right\vert e^{\frac{\tanh \lambda }{2}a^{2}}a^{\dag
n}e^{-\kappa ta^{\dagger }a}  \notag \\
& =\text{sech}\lambda \sum_{n=0}^{\infty }\frac{\left( T^{\prime }\tanh
\lambda \right) ^{n}}{2^{n}n!}e^{\frac{e^{-2\kappa t}a^{\dag 2}\tanh \lambda
}{2}}  \notag \\
& \times H_{n}\left( i\sqrt{\frac{\tanh \lambda }{2}}a^{\dagger }e^{-\kappa
t}\right) \left\vert 0\right\rangle \left\langle 0\right\vert  \notag \\
& \times H_{n}\left( -i\sqrt{\frac{\tanh \lambda }{2}}ae^{-\kappa t}\right)
e^{\frac{e^{-2\kappa t}a^{2}\tanh \lambda }{2}}  \notag \\
& =\text{sech}\lambda \sum_{n=0}^{\infty }\frac{\left( T^{\prime }\tanh
\lambda \right) ^{n}}{2^{n}n!}\colon e^{\frac{e^{-2\kappa t}\left(
a^{2}+a^{\dag 2}\right) \tanh \lambda }{2}-a^{\dagger }a}  \notag \\
& \times H_{n}\left( i\sqrt{\frac{\tanh \lambda }{2}}a^{\dagger }e^{-\kappa
t}\right) H_{n}\left( -i\sqrt{\frac{\tanh \lambda }{2}}ae^{-\kappa t}\right)
\colon  \label{p22}
\end{align}
then using the following identity \cite{16}
\begin{eqnarray}
&&\sum_{n=0}^{\infty }\frac{t^{n}}{2^{n}n!}H_{n}\left( x\right) H_{n}\left(
y\right)  \notag \\
&=&\left( 1-t^{2}\right) ^{-1/2}\exp \left[ \frac{t^{2}\left(
x^{2}+y^{2}\right) -2txy}{t^{2}-1}\right] ,  \label{p23}
\end{eqnarray}
and $e^{\lambda a^{\dag }a}=\colon e^{\left( e^{\lambda }-1\right) a^{\dag
}a}\colon ,$ we finally obtain the expression of the output state%
\begin{equation}
\rho \left( t\right) =We^{\frac{\text{\ss }}{2}a^{\dag 2}}e^{a^{\dagger
}a\ln \left( \text{\ss }T^{\prime }\tanh \lambda \right) }e^{\frac{\text{\ss
}}{2}a^{2}},  \label{p23a}
\end{equation}
with $T^{\prime }=1-e^{-2\kappa t}$ and
\begin{equation}
W\equiv \frac{\text{sech}\lambda }{\sqrt{1-T^{\prime 2}\tanh ^{2}\lambda }},
\text{\ss }\equiv \frac{e^{-2\kappa t}\tanh \lambda }{1-T^{\prime 2}\tanh
^{2}\lambda }.  \label{p24}
\end{equation}
By comparing Eq.(\ref{p15}) with (\ref{p23}) one can see that after going
through the channel the initial squeezing parameter $\tanh \lambda $ in Eq.(
\ref{p15}) becomes to \ss $\equiv \frac{e^{-2\kappa t}\tanh \lambda }{%
1-T^{\prime 2}\tanh ^{2}\lambda },$ and $\left\vert 0\right\rangle
\left\langle 0\right\vert \rightarrow \frac{1}{\sqrt{1-T^{\prime 2}\tanh
^{2}\lambda }}e^{a^{\dagger }a\ln \left( \text{\ss }T^{\prime }\tanh \lambda
\right) },$ a chaotic state (mixed state), due to $T^{\prime }>0,$ we can
prove $\frac{e^{-2\kappa t}}{1-T^{\prime 2}\tanh ^{2}\lambda }<1,$ which
means a squeezing-decreasing process. When $\kappa t=0$, then $T^{\prime }=0$
and \ss\ $=\tanh \lambda $, Eq.(\ref{p22}) becomes the initial squeezed
vacuum state as expected.

It is important to check: if Tr$\rho (t)=1$. Using Eq.(\ref{p22}) and the
completeness of coherent state $\int \frac{d^{2}z}{\pi }\left\vert
z\right\rangle \left\langle z\right\vert =1$ as well as the following
formula \cite{17}
\begin{equation}
\int \frac{d^{2}z}{\pi }e^{\zeta \left\vert z\right\vert ^{2}+\xi z+\eta
z^{\ast }+fz^{2}+gz^{\ast 2}}=\frac{1}{\sqrt{\zeta ^{2}-4fg}}e^{\frac{-\zeta
\xi \eta +f\eta ^{2}+g\xi ^{2}}{\zeta ^{2}-4fg}},  \label{p25}
\end{equation}%
whose convergent condition is Re$\left( \zeta \pm f\pm g\right) <0$ and$\
\mathtt{Re}\left( \frac{\zeta ^{2}-4fg}{\zeta \pm f\pm g}\right) <0$, we
really see
\begin{eqnarray}
\text{Tr}\rho \left( t\right) &=&W\int \frac{d^{2}z}{\pi }\left\langle
z\right\vert e^{\frac{\text{\ss }}{2}a^{\dag 2}}e^{a^{\dagger }a\ln \left(
\text{\ss }T^{\prime }\tanh \lambda \right) }e^{\frac{\text{\ss }}{2}
a^{2}}\left\vert z\right\rangle  \notag \\
&=&\frac{W}{\sqrt{\left( \text{\ss }T^{\prime }\tanh \lambda -1\right) ^{2}-
\text{\ss }^{2}}}=1.  \label{p26}
\end{eqnarray}
so $\rho \left( t\right) $ is qualified to be a mixed state, thus we see an
initial pure squeezed vacuum state evolves into a squeezed chaotic state
with decreasing-squeezing after passing through an amplitude dissipative
channel.

\section{Average photon number}

Using the completeness relation of coherent state and the normally ordering
form of $\rho \left( t\right) $ in Eq. (\ref{p22}), and using $e^{\frac{
\text{\ss }}{2}a^{2}}a^{\dagger }e^{-\frac{\text{\ss }}{2}a^{2}}=a^{\dagger
}+$\ss $a$, as well\ as $e^{a^{\dagger }a\ln \left( \text{\ss }T^{\prime
}\tanh \lambda \right) }a^{\dagger }e^{-a^{\dagger }a\ln \left( \text{\ss }
T^{\prime }\tanh \lambda \right) }$=$a^{\dagger }$\ss $T^{\prime }\tanh
\lambda ,$ we have
\begin{align}
& \mathtt{Tr}\left( \rho \left( t\right) a^{\dagger }a\right)  \notag \\
& =W\int \frac{d^{2}z}{\pi }\left\langle z\right\vert e^{\frac{\text{\ss }}{
2 }a^{\dag 2}}e^{a^{\dagger }a\ln \left( \text{\ss }T^{\prime }\tanh \lambda
\right) }e^{\frac{\text{\ss }}{2}a^{2}}a^{\dagger }a\left\vert z\right\rangle
\notag \\
& =W\int \frac{d^{2}z}{\pi }\left\langle z\right\vert e^{\frac{\text{\ss }}{
2 }a^{\dag 2}}e^{a^{\dagger }a\ln \left( \text{\ss }T^{\prime }\tanh \lambda
\right) }e^{\frac{\text{\ss }}{2}a^{2}}za^{\dagger }\left\vert z\right\rangle
\notag \\
& =W\int \frac{d^{2}z}{\pi }z\left\langle z\right\vert e^{\frac{\text{\ss }}{
2}a^{\dag 2}}e^{a^{\dagger }a\ln \left( \text{\ss }T^{\prime }\tanh \lambda
\right) }\left( a^{\dagger }+\text{\ss }a\right) e^{\frac{\text{\ss }}{2}
a^{2}}\left\vert z\right\rangle  \notag \\
& =W\text{\ss }\int \frac{d^{2}z}{\pi }ze^{\frac{\text{\ss }}{2}\left(
z^{\ast 2}+z^{2}\right) }\left\langle z\right\vert \left( a^{\dagger
}T^{\prime }\tanh \lambda +z\right) e^{a^{\dagger }a\ln \left( \text{\ss }
T^{\prime }\tanh \lambda \right) }\left\vert z\right\rangle  \notag \\
& =W\text{\ss }\int \frac{d^{2}z}{\pi }\left( |z|^{2}T^{\prime }\tanh
\lambda +z^{2}\right)  \notag \\
& \times \exp \left[ \left( \text{\ss }T^{\prime }\tanh \lambda -1\right)
|z|^{2}+\frac{\text{\ss }}{2}\left( z^{\ast 2}+z^{2}\right) \right] .
\label{p59}
\end{align}
In order to perform the integration, we reform Eq.(\ref{p59}) as $
\allowbreak $
\begin{eqnarray}
\mathtt{Tr}\left( \rho \left( t\right) a^{\dagger }a\right) &=&W\text{\ss }
\left\{ T^{\prime }\tanh \lambda \frac{\partial }{\partial f}+\frac{2}{\text{
\ss }}\frac{\partial }{\partial s}\right\}  \notag \\
&&\times \int \frac{d^{2}z}{\pi }\exp \left[ \left( \text{\ss }T^{\prime
}\tanh \lambda -1+f\right) |z|^{2}\right.  \notag \\
&&+\left. \frac{\text{\ss }}{2}\left( z^{\ast 2}+\left( 1+s\right)
z^{2}\right) \right] _{f=s=0}  \notag \\
&=&\frac{1-\text{\ss }T^{\prime }\tanh \lambda }{\left( \text{\ss }T^{\prime
}\tanh \lambda -1\right) ^{2}-\text{\ss }^{2}}-1  \label{p27}
\end{eqnarray}
in the last step, we have used Eq.(\ref{p26}). Using Eq.(\ref{p27}), we
present the time evolution of the average photon number in Fig. 1, from
which we find that the average photon number of the single-mode squeezed
vacuum state in the amplitude damping channel reduces gradually to zero when
decay time goes.
\begin{figure}[tbp]
\centering \includegraphics[width=8cm]{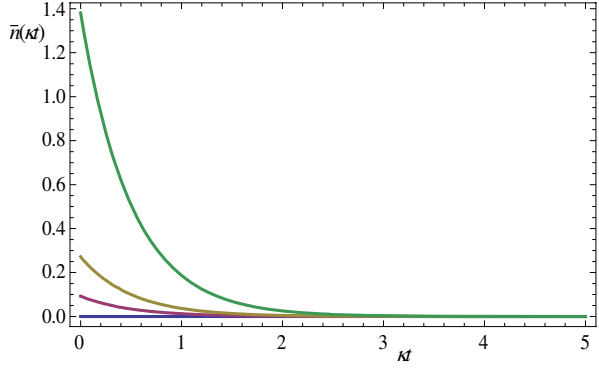}
\caption{(Color online) The average $\bar{n}\left( \protect\kappa t\right) $
as the function of $\protect\kappa t$ for different values of squeezing
parameter $\protect\lambda $ (from bottom $\ $\ to top $\protect\lambda
=0,0.1,0.3,0.5,1$.) }
\end{figure}

\section{Photon statistics distribution}

Next, we shall derive the photon statistics distributions of $\rho \left(
t\right) $. The photon number is given by $p\left( n,t\right) =\left\langle
n\right\vert \rho \left( t\right) \left\vert n\right\rangle $. Noticing $
a^{\dag m}\left\vert n\right\rangle =\sqrt{(m+n)!/n!}\left\vert
m+n\right\rangle $ and using the un-normalized coherent state $\left\vert
\alpha \right\rangle =\exp [\alpha a^{\dag }]\left\vert 0\right\rangle $,
\cite{18,19} leading to $\left\vert n\right\rangle =\frac{1}{\sqrt{n!}}\frac{
\mathtt{d}^{n}}{\mathtt{d}\alpha ^{n}}\left\vert \alpha \right\rangle
\left\vert _{\alpha =0}\right. ,$ $\left( \left\langle \beta \right.
\left\vert \alpha \right\rangle =e^{\alpha \beta ^{\ast }}\right) $, as well
as the normal ordering form of $\rho \left( t\right) $ in Eq. (\ref{p22}),
the probability of finding $n$ photons in the field is given by
\begin{eqnarray}
&&p\left( n,t\right)  \notag \\
&=&\left\langle n\right\vert \rho \left( t\right) \left\vert n\right\rangle
\notag \\
&=&\frac{W}{n!}\frac{\mathtt{d}^{n}}{\mathtt{d}\beta ^{\ast n}}\frac{\mathtt{
\ d}^{n}}{\mathtt{d}\alpha ^{n}}\left. \left\langle \beta \right\vert e^{
\frac{ \text{\ss }}{2}\beta ^{\ast 2}}e^{a^{\dagger }a\ln \left( \text{\ss }
T^{\prime }\tanh \lambda \right) }e^{\frac{\text{\ss }}{2}\alpha
^{2}}\left\vert \alpha \right\rangle \right\vert _{\alpha ,\beta ^{\ast }=0}
\notag \\
&=&\frac{W}{n!}\frac{\mathtt{d}^{n}}{\mathtt{d}\beta ^{\ast n}}\frac{\mathtt{
\ d}^{n}}{\mathtt{d}\alpha ^{n}}\left. \exp \left[ \beta ^{\ast }\alpha
\text{ \ss }T^{\prime }\tanh \lambda +\frac{\text{\ss }}{2}\beta ^{\ast 2}+
\frac{ \text{\ss }}{2}\alpha ^{2}\right] \right\vert _{\alpha ,\beta ^{\ast
}=0}.  \label{p49}
\end{eqnarray}
Note that
\begin{equation*}
\left[ e^{\frac{\text{\ss }}{2}a^{\dagger 2}}e^{a^{\dagger }a\ln \left(
\text{\ss }T^{\prime }\tanh \lambda \right) }e^{\frac{\text{\ss }}{2}\alpha
^{2}}\right] ^{\dagger }=e^{\frac{\text{\ss }}{2}a^{\dagger 2}}e^{a^{\dagger
}a\ln \left( \text{\ss }T^{\prime }\tanh \lambda \right) }e^{\frac{\text{\ss
}}{2}\alpha ^{2}}
\end{equation*}
so
\begin{equation*}
\left\langle n\right\vert \rho \left( t\right) \left\vert n\right\rangle
^{\ast }=\left\langle n\right\vert \rho \left( t\right) ^{\dagger
}\left\vert n\right\rangle =\left\langle n\right\vert \rho \left( t\right)
\left\vert n\right\rangle
\end{equation*}
\begin{eqnarray}
&&\frac{\partial ^{n+n}}{\partial t^{n}\partial t^{\prime n}}\exp \left[
2xtt^{\prime }-t^{2}-t^{\prime 2}\right] _{t=t^{\prime }=0}  \notag \\
&=&2^{n}n!\sum_{m=0}^{[n/2]}\frac{n!}{2^{2m}\left( m!\right) ^{2}(n-2m)!}
x^{n-2m},  \label{p50}
\end{eqnarray}
we derive the compact form for $\mathfrak{p}\left( n,t\right) $, i.e.,\
\begin{eqnarray}
&&p\left( n,t\right)  \notag \\
&=&\frac{W}{n!}\left( -\frac{\text{\ss }}{2}\right) ^{n}\frac{\mathtt{d}^{n}
}{\mathtt{d}\beta ^{\ast n}}\frac{\mathtt{d}^{n}}{\mathtt{d}\alpha ^{n}}
\left. e^{-2T^{\prime }\tanh \lambda \beta ^{\ast }\alpha -\beta ^{\ast
2}-\alpha ^{2}}\right\vert _{\alpha ,\beta ^{\ast }=0}  \notag \\
&=&W\left( \text{\ss }T^{\prime }\tanh \lambda \right) ^{n}\sum_{m=0}^{[n/2]}
\frac{n!\left( T^{\prime }\tanh \lambda \right) ^{-2m}}{2^{2m}\left(
m!\right) ^{2}(n-2m)!}.  \label{p52}
\end{eqnarray}%
Using the newly expression of Legendre polynomials found in Ref. \cite{20}
\begin{equation}
x^{n}\sum_{m=0}^{[n/2]}\frac{n!}{2^{2m}\left( m!\right) ^{2}(n-2m)!}\left( 1-
\frac{1}{x^{2}}\right) ^{m}=P_{n}\left( x\right) ,  \label{p51}
\end{equation}
we can formally recast Eq.(\ref{p52}) into the following compact form, i.e.,
\begin{equation*}
p\left( n,t\right) =W\left( e^{-\kappa t}\sqrt{-\text{\ss }\tanh \lambda }
\right) ^{n}P_{n}\left( e^{\kappa t}T^{\prime }\sqrt{-\text{\ss }\tanh
\lambda }\right)
\end{equation*}
note that since $\sqrt{-\text{\ss }\tanh \lambda }$ is pure imaginary, while
$p\left( n,t\right) $ is real, so we must still use the power-series
expansion on the right-hand side of Eq.(\ref{p52}) to depict figures of the
variation of $p\left( n,t\right) $. In particular, when $t=0$, Eq.(\ref{p52}
) reduces to
\begin{eqnarray}
p\left( n,0\right) &=&\text{sech}\lambda \left( \tanh \lambda \right)
^{n}\lim_{T^{\prime }\rightarrow 0}\sum_{m=0}^{[n/2]}\frac{n!\left(
T^{\prime }\tanh \lambda \right) ^{n-2m}}{2^{2m}\left( m!\right) ^{2}(n-2m)!}
\notag \\
&=&\left\{
\begin{array}{cc}
\frac{\left( 2k\right) !}{2^{2k}k!k!}\text{sech}\lambda \tanh ^{2k}\lambda ,
& n=2k \\
0 & n=2k+1
\end{array}
\right. ,  \label{p53}
\end{eqnarray}
which just correspond to the number distributions of the squeezed vacuum
state \cite{21,22}. From\ Eq.(\ref{p53}) it is not difficult to see that the
photocount distribution decreases as the squeezing parameter $\lambda $
increases. While for $\kappa t\rightarrow \infty ,$ we see that $p\left(
n,\infty \right) =0.$ This indicates that there is no photon when a system
interacting with a amplitude dissipative channel for enough long time, as
expected. In Fig. 2, the photon number distribution is shown for different $
\kappa t$.
\begin{figure}[t]
\label{2}\centering\includegraphics[width=8cm]{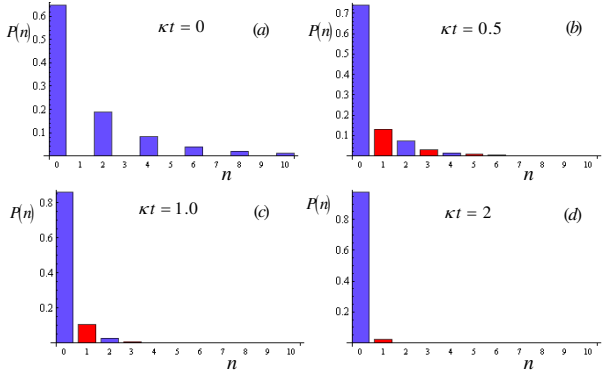}
\caption{(Color online) Photon number distribution of the squeezed vacuum
state in amplitude damping channel for $\protect\lambda =1$, and different $
\protect\kappa t$: ($a)$ $\protect\kappa t=0$, ($b$) $\protect\kappa t=0.5,$
($c$) $\protect\kappa t=1$ and ($d$) $\protect\kappa t=2$.}
\end{figure}

\section{Wigner functions}

In this section, we shall use the normally ordering for of density operators
to calculate the analytical expression of Wigner function. For a single-mode
system, the WF is given by \cite{23}
\begin{equation}
W\left( \alpha ,\alpha ^{\ast },t\right) =e^{2\left\vert \alpha \right\vert
^{2}}\int \frac{d^{2}\beta }{\pi ^{2}}\left\langle -\beta \right\vert \rho
\left( t\right) \left\vert \beta \right\rangle e^{-2\left( \beta \alpha
^{\ast }-\beta ^{\ast }\alpha \right) },  \label{p60}
\end{equation}
where $\left\vert \beta \right\rangle $ is the coherent state \cite{18,19} .
From Eq.(\ref{p22}) it is easy to see that once the normal ordered form of $
\rho \left( t\right) $ is known, we can conveniently obtain the Wigner
function of $\rho \left( t\right) $.

On substituting Eq.(\ref{p23a}) into Eq.(\ref{p60}) we obtain the WF of the
single-mode squeezed state in the ADC,
\begin{eqnarray}
&&W\left( \alpha ,\alpha ^{\ast },t\right)  \notag \\
&=&We^{2\left\vert \alpha \right\vert ^{2}}\int \frac{d^{2}\beta }{\pi ^{2}}
\exp \left[ -\left( 1+\text{\ss }T^{\prime }\tanh \lambda \right) \left\vert
\beta \right\vert ^{2}\right.  \notag \\
&&\left. -2\left( \beta \alpha ^{\ast }-\beta ^{\ast }\alpha \right) +\frac{
\text{\ss }}{2}\beta ^{\ast 2}+\frac{\text{\ss }}{2}\beta ^{2}\right]  \notag
\\
&=&\frac{W}{\pi \sqrt{\left( 1+\text{\ss }T^{\prime }\tanh \lambda \right)
^{2}-\text{\ss }^{2}}}\exp \left[ 2\left\vert \alpha \right\vert ^{2}\right]
\notag \\
&&\times \exp \left[ 2\frac{-2\left( 1+\text{\ss }T^{\prime }\tanh \lambda
\right) \left\vert \alpha \right\vert ^{2}+\text{\ss }\left( \alpha ^{\ast
2}+\alpha ^{2}\right) }{\left( 1+\text{\ss }T^{\prime }\tanh \lambda \right)
^{2}-\text{\ss }^{2}}\right]  \label{p61}
\end{eqnarray}
In particular, when $t=0$ and $t\rightarrow \infty $, Eq.(\ref{p61}) reduces
to $W\left( \alpha ,\alpha ^{\ast },0\right) =\frac{1}{\pi }\exp
[-2\left\vert \alpha \right\vert ^{2}\cosh 2\lambda +\left( \alpha ^{\ast
2}+\alpha ^{2}\right) \sinh 2\lambda ]$, and $W\left( \alpha ,\alpha ^{\ast
},\infty \right) =\frac{1}{\pi }\exp \left[ -2\left\vert \alpha \right\vert
^{2}\right] $, which are just the WF of the single-mode squeezed vacuum
state and the vacuum state, respectively. In Fig. 3, the WF of the
single-mode squeezed vacuum state in the amplitude damping channel is shown
for different decay time $\kappa t$.
\begin{figure}[t]
\centerline{\includegraphics[width=8cm]{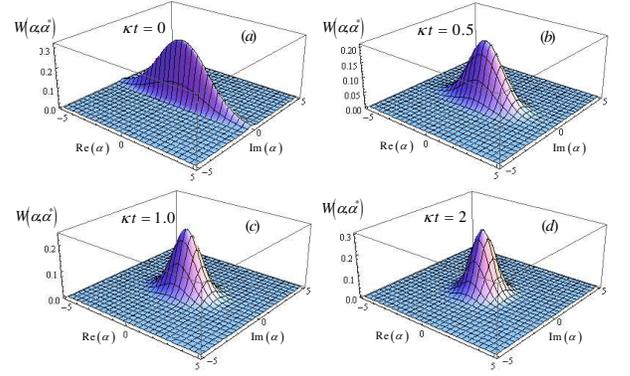}}
\caption{(Color online) Wigner function of the squeezed vacuum state in
amplitude damping channel for $\protect\lambda =1.0$, different $\protect
\kappa t$: ($a)$ $\protect\kappa t=0.0$, ($b$) $\protect\kappa t=0.5$, ($c$)
$\protect\kappa t=1$, and ($d$) $\protect\kappa t=2$.}
\end{figure}

\section{Tomogram}

As we know, once the probability distributions $P_{\theta }\left( \hat{x}
_{\theta }\right) $ of the quadrature amplitude are obtained, one can use
the inverse Radon transformation familiar in tomographic imaging to obtain
the WF and density matrix \cite{24}. Thus the Radon transform of the WF is
corresponding to the probability distributions $P_{\theta }\left( \hat{x}
_{\theta }\right) $. In this section we derive the tomogram of $\rho \left(
t\right) $.

For a single-mode system, the Radon transform of WF, denoted as $\mathcal{R}$
is defined by \cite{25}
\begin{eqnarray}
\mathcal{R}\left( q\right) _{f,g} &=&\int \delta \left( q-fq^{\prime
}-gp^{\prime }\right) Tr\left[ \Delta \left( \beta \right) \rho \left(
t\right) \right] dq^{\prime }dp^{\prime }  \notag \\
&=&Tr\left[ \left \vert q\right \rangle _{f,g\text{ }f,g}\left \langle
q\right \vert \rho \left( t\right) \right] =_{f,g}\left \langle q\right
\vert \rho \left( t\right) \left \vert q\right \rangle _{f,g}  \label{p63}
\end{eqnarray}
where the operator $\left \vert q\right \rangle _{f,g\text{ }
f,g}\left
\langle q\right \vert $ is just the Radon transform of
single-mode Wigner operator $\Delta \left( \beta \right) $, and
\begin{equation}
\left \vert q\right \rangle _{f,g}=A\exp \left[ \frac{\sqrt{2}qa^{\dag }}{B}
- \frac{B^{\ast }}{2B}a^{\dag 2}\right] \left \vert 0\right \rangle ,
\label{p64}
\end{equation}
as well as $B=f-ig,$ $A=\left[ \pi \left( f^{2}+g^{2}\right) \right]
^{-1/4}\exp [-q^{2}/2\left( f^{2}+g^{2}\right) ]$. Thus the tomogram of a
quantum state $\rho \left( t\right) $ is just the quantum average of $\rho
\left( t\right) $ in $\left \vert q\right \rangle _{f,g}$ representation (a
kind of intermediate coordinate-momentum representation) \cite{26}.

Substituting Eqs.(\ref{p23a}) and (\ref{p64}) into Eq.(\ref{p63}), and using
the completeness relation of coherent state, we see that the Radom transform
of WF of $\rho \left( t\right) $ is given by
\begin{eqnarray}
&&\mathcal{R}\left( q\right) _{f,g}  \notag \\
&=&W_{f,g}\left\langle q\right\vert e^{\frac{\text{\ss }}{2}a^{\dag
2}}e^{a^{\dagger }a\ln \left( \text{\ss }T^{\prime }\tanh \lambda \right)
}e^{\frac{\text{\ss }}{2}a^{2}}\left\vert q\right\rangle _{f,g}  \notag \\
&=&\frac{WA^{2}}{\sqrt{E}}\exp \left\{ \frac{q^{2}\text{\ss }}{E\left\vert
B\right\vert ^{4}}\left( B^{2}+B^{\ast }{}^{2}\right) \right.  \notag \\
&&+\left. \frac{2q^{2}\text{\ss }}{E\left\vert B\right\vert ^{2}}\left(
T\tanh \lambda +\text{\ss }-\text{\ss }T^{2}\tanh ^{2}\lambda \right)
\right\} ,  \label{p65}
\end{eqnarray}
where we have used the formula (\ref{p25}) and $\left\langle \alpha
\right\vert \left. \gamma \right\rangle =\exp [-\left\vert \alpha
\right\vert ^{2}/2-\left\vert \gamma \right\vert ^{2}/2+\alpha ^{\ast
}\gamma ]$, as well as$\allowbreak $
\begin{eqnarray}
E &=&\left( 1+\text{\ss }\frac{B}{B^{\ast }}\right) \left( 1+\frac{B^{\ast }
}{B}\text{\ss }-B^{\ast }\frac{\left( \text{\ss }T^{\prime }\tanh \lambda
\right) ^{2}}{B^{\ast }+\text{\ss }B}\right)  \notag \\
&=&\left\vert 1+\frac{\text{\ss }B}{B^{\ast }}\right\vert ^{2}-\left( \text{
\ss }T^{\prime }\tanh \lambda \right) ^{2}.  \label{p66}
\end{eqnarray}
In particular, when $t=0,$ ($T=0$), then Eq.(\ref{p65}) reduces to ($\frac{B
}{B^{\ast }}=e^{2i\phi }$)
\begin{eqnarray}
\mathcal{R}\left( q\right) _{f,g} &=&\frac{A^{2}\text{sech}\lambda }{
\left\vert 1+e^{2i\phi }\tanh \lambda \right\vert }  \notag \\
&&\times \exp \left\{ \frac{q^{2}\left( B^{2}+B^{\ast }{}^{2}+2\left\vert
B\right\vert ^{2}\tanh \lambda \right) \tanh \lambda }{\left\vert
1+e^{2i\phi }\left\vert B\right\vert ^{4}\tanh \lambda \right\vert ^{2}}
\right\} ,  \label{p67}
\end{eqnarray}
which is a tomogram of single-mode squeezed vacuum state; while for $\kappa
t\rightarrow \infty ,$($T=1$), then $\mathcal{R}\left( q\right)
_{f,g}=A^{2}, $ which is a Gaussian distribution corresponding to the vacuum
state.

In summary, using the way of deriving infinitive sum representation of
density operator by virtue of the entangled state representation describing,
we conclude that in the amplitude dissipative channel the initial density
operator of a single-mode squeezed vacuum state evolves into a squeezed
chaotic state with decreasing-squeezing. We investigate average photon
number, photon statistics distributions, Wigner functions and tomogram for
the output state.

\section*{Acknowledgments}

This work is supported by the National Natural Science Foundation of China
(Grant No.11175113 and 11047133), Shandong Provincial Natural Science
Foundation in China (Gant No.ZR2010AQ024), and a grant from the Key Programs
Foundation of Ministry of Education of China (Grant No. 210115),as well as
Jiangxi Provincial Natural Science Foundation in China (No. 2010GQW0027).

\end{document}